\documentclass[twocolumn,showpacs,amsmath,floatfix,prb,aps]{revtex4}

\usepackage{color}
\usepackage{amsmath}
\usepackage{pifont}   % Ding symbols
\usepackage{graphicx} % Include figure files
\usepackage{dcolumn}  % Align table columns on decimal point
\usepackage{bm}       % bold math
\usepackage{amsfonts} % Some more fonts
\usepackage{amssymb}  % More symbols
\usepackage{multirow} % Table functions
\usepackage{siunitx}

\begin{document}
%\DeclareGraphicsRule{*}{png}{*}{}

%%%% User-defined commands %%%%
\newcommand{\ba}{{\bf a}}
\newcommand{\BB}{{\bf b}}
\newcommand{\bbb}{{\bf B}}
\newcommand{\bd}{{\bf d}}
\newcommand{\br}{{\bf r}}
\newcommand{\bp}{{\bf p}}
\newcommand{\bk}{{\bf k}}
\newcommand{\bg}{{\bf g}}
\newcommand{\bj}{{\bf j}}
\newcommand{\bt}{{\bf t}}
\newcommand{\bv}{{\bf v}}
\newcommand{\bu}{{\bf u}}
\newcommand{\bq}{{\bf q}}
\newcommand{\bG}{{\bf G}}
\newcommand{\bP}{{\bf P}}
\newcommand{\bJ}{{\bf J}}
\newcommand{\bK}{{\bf K}}
\newcommand{\bL}{{\bf L}}
\newcommand{\bR}{{\bf R}}
\newcommand{\bS}{{\bf S}}
\newcommand{\bT}{{\bf T}}
\newcommand{\bQ}{{\bf Q}}
\newcommand{\bA}{{\bf A}}
\newcommand{\bH}{{\bf H}}

\newcommand{\bra}[1]{\left\langle #1 \right |}
\newcommand{\ket}[1]{\left| #1 \right\rangle}
\newcommand{\braket}[2]{\left\langle #1 | #2 \right\rangle}
\newcommand{\mel}[3]{\left\langle #1 \left| #2 \right| #3 \right\rangle}

\newcommand{\bdel}{\boldsymbol{\delta}}
\newcommand{\bsig}{\boldsymbol{\sigma}}
\newcommand{\beps}{\boldsymbol{\epsilon}}
\newcommand{\bnu}{\boldsymbol{\nu}}
\newcommand{\bnab}{\boldsymbol{\nabla}}
\newcommand{\bGam}{\boldsymbol{\Gamma}}

\newcommand{\bgt}{\tilde{\bf g}}

\newcommand{\brh}{\hat{\bf r}}
\newcommand{\bph}{\hat{\bf p}}

\author{D. Weckbecker$^1$}
\author{R. Gupta$^1$}
\author{F. Rost$^1$}
\author{S. Sharma$^2$}
\author{S. Shallcross$^1$}
\email{sam.shallcross@fau.de}
\affiliation{1 Lehrstuhl f\"ur Theoretische Festk\"orperphysik, Staudtstr. 7-B2, 91058 Erlangen, Germany.}
\affiliation{2 Max-Planck-Institut fur Mikrostrukturphysik Weinberg 2, D-06120 Halle, Germany.}

\title{Dislocation and node states in bilayer graphene systems}
\date{\today}

\begin{abstract}
We investigate the electronic structure of realistic partial dislocation networks in bilayer graphene that feature annihilating, wandering, and intersecting partial lines. We find charge accumulation states at partials that are sensitive to Fermi energy and partial Burgers vector but not to the screw versus edge character of the partial. 
These states are shown to be current carrying, with the current density executing a spiral motion along the dislocation line with a strong interlayer component to the current. Close to the Dirac point localization on partials switches to localization on intersections of partials, with a corresponding complex current flow around the nodes. 
%numbers?
%edge state?
\end{abstract}

%\pacs{73.20.At, 73.21.Ac, 81.05.Uw}

\maketitle

\section{Introduction}

Bernal stacked bilayer graphene is an imperfect material. Transmission electron microscopy (TEM) experiments have revealed the existence of a network of partial dislocations extending throughout the material\cite{ald13,butz14}, which has been shown to impact electronic properties dramatically in further investigations\cite{kiss15,2,3,8,12,4,6,9,shall17}. In fact, even a single partial dislocation can qualitatively change transport at the Dirac point, driving a nominally minimally metallic sample to an insulating state\cite{shall17}. Alternatively, in a semi-conducting bilayer, partial dislocations provide a network of conducting channels through topologically protected bound states\cite{4,6}, creating a metal out of a semi-conductor. Such extended defects arising from the stacking degree of freedom are likely to be found in all van der Walls bound multilayers -- a broad class of materials. Bilayer graphene, probably the simplest such material, thus offers an ideal test ground for elucidating the impact of this important defect on low-dimensional systems.

However, even for bilayer graphene the electronic structure of partial dislocations is not fully understood, a situation attributable to the formidable computation challenge posed by simulating a realistic dislocation network. 
For example, in graphene grown epitaxially on SiC the domain size is of the order of $50\,$nm and the partial width of the order of $5\,$nm. 
A network of $\sim20$ domains then requires of the order of a square micrometer, equivalent to $\sim10^8$ carbon atoms. This lies well beyond the scope of any atomistic approach and taxes even continuum methodologies. 
To date, therefore, most theoretical studies have considered idealized partial dislocations that are (i) straight and (ii) do not intersect with other partial dislocations. The purpose of the present paper is to address the electronic structure of realistic partial dislocation networks in which partials both wander and intersect.

We consider two networks, one derived directly from experimental TEM images -- of the order of 1$\mu m^2$ in area -- and a model network of trigonally intersecting partials 100 times smaller in area. Our principal focus will be the question of localized states on partial dislocations and their intersections. While in a finite layer-perpendicular electric field bound states on partials are topologically protected, there is evidence that they exist even in the absence of such a field\cite{8,12}. We find such localized states, and show that their existence is sensitive to the partial Burgers vector and energy, although not to the screw versus edge character of a wandering partial. At energies below the Dirac point so-called type 3 partials feature localized states, switching to type 2 partials above the Dirac point, with close to the Dirac point localization instead occurring on intersections of the network. This coupling of energy and Burgers vector is a robust feature found in both the experimental as well as model network.

We further analyze the current density of partial dislocation networks, finding intense currents both along partials and at nodes, with a strong interlayer component. For example, the localized states on partials of type 2 and 3 are shown to possess a current density spiraling around the dislocation line, with the spiral chirality opposite in opposite K valleys. In general, we find that a current density impinging upon a partial is driven from one layer onto the other and then back again; a truncated version of the ``snake states'' found in twist bilayer graphene\cite{mel10,shall10,san12,shall13,shall16}. On the other hand, for the node states found at energies close to the Dirac point we find strong current densities around the vertices of intersecting partials, with a complex pattern of in-plane and interlayer currents.

\section{Theory}

\subsection{Continuum description of a 
partial network}\label{Sfield}

A partial dislocation in bilayer graphene represents a continuous change in stacking order from AB to AC stacking. Their electronic description therefore requires a stacking potential dependent on the local displacement of the layers at each point of the lattice. Taking AB stacking as the reference state this potential will, in the low-energy Dirac-Weyl sector, be represented by a $2\times2$ matrix $S(\br)$ in a Hamiltonian of the form:

\begin{equation}
 H(\br,\bp) = \begin{pmatrix}
      v_F \bsig.\bp & S(\br) \\
      S(\br)^\dagger & v_F \bsig^\ast.\bp
     \end{pmatrix}
     \label{H}
\end{equation}
such that $S(\br)$ couples the two Dirac-Weyl operators of each layer.
Such an interlayer stacking potential has been derived in Ref.~\onlinecite{M} and previously employed to treat dislocations in Refs.~\onlinecite{kiss15} and \onlinecite{shall17}. It takes the form

\begin{equation}
 S(\br) = \frac{1}{V_{UC}} \sum_i M_i \,\, t_\perp(K_i^2) e^{i\bG_i.\Delta \bu(\br)} \, ,
\label{S}
\end{equation}
where $\Delta \bu(\br) = \bu_2(\br) - \bu_1(\br)$ is the local displacement of the two layers. In this expression the $M_i$ matrices are given in Table \ref{M2}, the sum is over the translation group of the high-symmetry K point $\bK_1=\frac{4\pi}{3a} (1,0)$ with $a$ the lattice constant of graphene (\emph{i.e.}\ $\bK_i = \bK_1 + \bG_i$ with $\bG_i$ a reciprocal vector), and $t_\perp(q^2)$ is the Fourier transform of the interlayer hopping $t_\perp(\delta^2)$. From Eq.~\eqref{H} we see that the potential $S(\br)$ is a matrix whose elements describe the scattering amplitude from sublattice $\alpha$ of layer 1 to sublattice $\beta$ of layer 2; in the Dirac-Weyl language $S(\br)$ couples the pseudospin channels of each layer.

\begin{table}
\begin{tabular}{l|ccc}	
Stacking & $M_1$ & $M_2$ & $M_3$ \\ \hline\hline
 AB &
 $\begin{pmatrix} 1 & 1  \\ 1 & 1 \end{pmatrix}$ &
 $\begin{pmatrix} 1 & e^{i2\pi/3}  \\ e^{i2\pi/3} & e^{-i2\pi/3} \end{pmatrix}$ &
 $\begin{pmatrix} 1 & e^{-i2\pi/3}  \\ e^{-i2\pi/3} & e^{i2\pi/3} \end{pmatrix}$ \\
\hline
 AC &
 $\begin{pmatrix} 1 & 1  \\ 1 & 1 \end{pmatrix}$ &
 $\begin{pmatrix} e^{i2\pi/3} & e^{-i2\pi/3}  \\ e^{-i2\pi/3} & 1 \end{pmatrix}$ &
 $\begin{pmatrix} e^{-i2\pi/3} & e^{i2\pi/3}  \\ e^{i2\pi/3} & 1 \end{pmatrix}$
\end{tabular}
\caption{$M$ matrices for the $C_3$ star of Bernal stacked AB and AC stacked bilayer graphene.}
\label{M2}
\end{table}

\begin{figure}[tbp]
  \centering
  \includegraphics[width=0.98\linewidth]{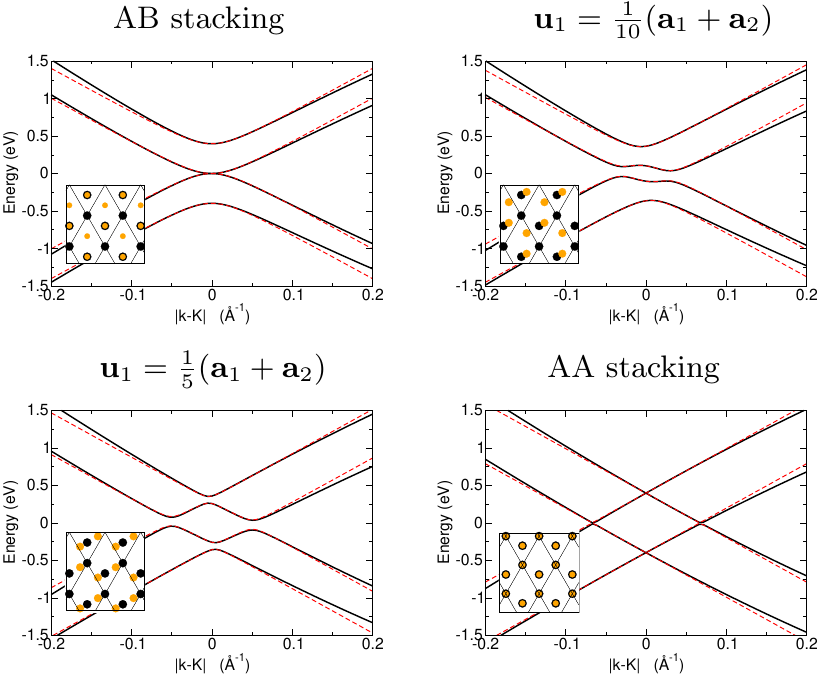}
  \caption{Low-energy electronic structure of bilayer graphene for four different mutual translations of the two layers that take the stacking from AB to AA type: (a) $\bu_1 = {\bf0}$ (AB stacking), (b) $\bu_1 = 1/10(\ba_1+\ba_2)$, (c) $\bu_1 = 1/5(\ba_1+\ba_2)$, (d) $\bu_1 = 1/3(\ba_1+\ba_2)$ (AA stacking). The full (black) lines are the result of a tight binding calculation, while the dashed (red) lines display results obtained from the interlayer stacking potential, Eq.~\eqref{S}, in conjunction with the Dirac-Weyl equation. The inset in each picture displays the lattice structure of the bilayer.
  }
  \label{shift}
\end{figure}

Note that the in-plane deformation accompanying a local shift of the two layers is not included in Eq.~\eqref{H}. The impact of such in-plane deformation upon the electronic structure of partial dislocations has been explored in Ref.~\onlinecite{shall17} and found to be negligible, and so we do not consider the resulting gauge, scalar, and Fermi velocity correction terms in the layer-diagonal blocks of our effective Hamiltonian\cite{gupta18}.

In order to establish that this approach can adequately describe the electronic properties of arbitrarily stacked bilayer graphene, we first analyze a very simple test case: a constant global shift of one layer. For a constant shift $\mathbf{u}_1$ applied to the first layer the stacking potential in Eq.~\eqref{S} becomes a constant matrix, with the effective Hamiltonian now being
\begin{equation}
 H(\bp,\bu_1) = \begin{pmatrix} v_F \bsig.\bp & S(\bu_1) \\ S(\bu_1)^\dagger & v_F \bsig^\ast.\bp \end{pmatrix}
 \label{HU} \, ,
\end{equation}
which may be directly diagonalized. In Fig.~\ref{shift} we show the low-energy spectrum of the bilayer for four shift vectors on a path that translates the bilayer from AB to AA stacking: $\bu_1 = {\bf 0}$ (the AB stacked bilayer), $\bu_1 = 1/10(\ba_1+\ba_2)$, $\bu_1 = 1/5(\ba_1+\ba_2)$, and $\bu_1 = 1/3(\ba_1+\ba_2)$ (AA stacked). Tight-binding results are shown by the full (black) lines with the results of the effective Hamiltonian presented with broken (red) lines. As may be seen, an excellent agreement exists between the two methods for (the relevant) energies of $\sim1\,$eV around the Dirac point, and so Eq.~\eqref{HU} captures the low-energy electronic structure for all mutual translations of the bilayer.

\begin{figure}
  \centering
  \includegraphics[width=0.85\linewidth]{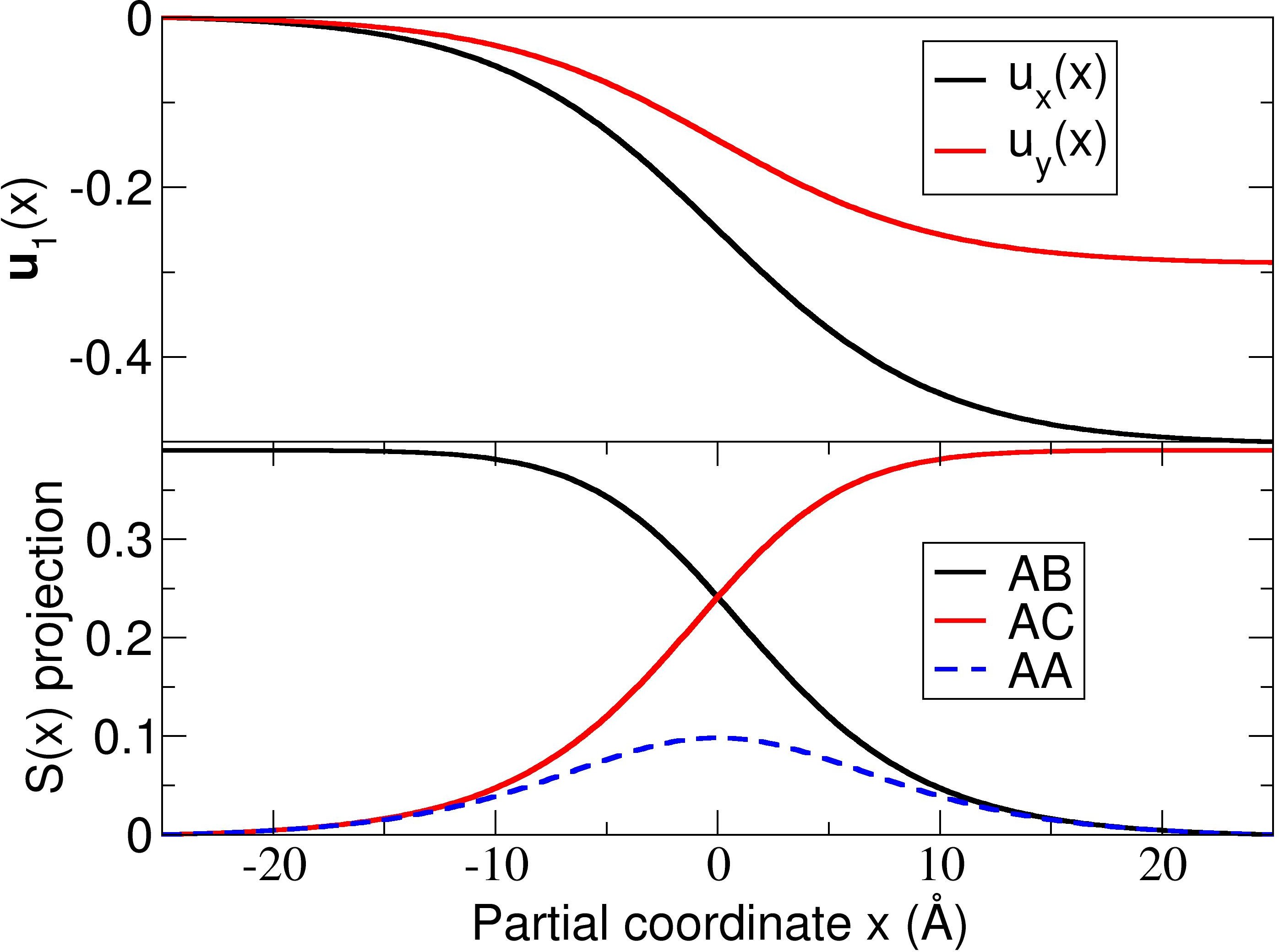}
  \caption{\emph{Upper panel}: Deformation field $\bu(x)$ applied to layer one of an AB stack bilayer to create a partial dislocation smoothly connecting regions of AB and AC stacking; $x$ is a coordinate perpendicular to the partial. On the left hand side $\bu = {\bf 0}$ (AB stacking) and on the right hand side $\bu = \bbb_3 = a(-1/2,-1/(2\sqrt{3}))$ (AC stacking). \emph{Lower panel}: The corresponding matrix valued effective interlayer field $S(x)$ required to describe the electronic structure of this partial dislocation. This is shown as a projection onto the three high-symmetry stacking types of the graphene bilayer AB, AC, and AA stacking (see Eqs.~\eqref{OH}-\eqref{yesss} for the projection matrices), showing clearly the transition from AB to AC stacking but also that the maximum component of AA stacking is found at the core of the partial.
  }
  \label{uphi}
\end{figure}

The requirement of a continuous graphene membrane in each layer leads to the condition that separate AB and AC domains be connected by one of three possible partial Burgers vectors: $\bbb_1 = a(1/2,1/(2\sqrt{3}))$, $\bbb_2 = a(-1/2,1/(2\sqrt{3}))$, and $\bbb_3=a(0,-1/\sqrt{3})$. Traversing from one domain to the other then involves a local shift of one layer by one of these partial Burgers vectors. This transition occurs, according to experiment, over a width of $\sim5\,$nm and the detailed atomic structure of this transition region has been carefully investigated via semi-empirical bond potential calculations in Ref.~\onlinecite{butz14}. From this data we are able to extract a model form of $\bu_1(\br)$,

\begin{equation}
 \bu_1 = \frac{\bbb_i}{2}\left(1+\frac{\tanh\gamma x}{\tanh\gamma L_0/2}\right)
 \label{disp}
\end{equation}
%
% values of coefficients?
with $\bbb_i$ the partial Burgers vector of the partial dislocation, $L_0$ the partial width, and $\gamma$ the partial stiffness, and the values of $L_0$ and $\gamma$ chosen to reproduce closely the data of Ref.~\onlinecite{butz14}. This is shown in Fig.~\ref{uphi} for the $AB \to AC$ transition mediated by $\bbb_3$ partial Burgers vector. Also shown is the interlayer field $S(\br)$, projected onto the 3 distinct high-symmetry stacking types that exist through this transition, $S_{AB}$, $S_{AC}$, and $S_{AA}$:

\begin{eqnarray}
 S_{AB} & = & \begin{pmatrix} 1 & 0 \\ 0 & 0 \end{pmatrix}, \label{OH}\\
 S_{AC} & = & e^{-i2\pi/3}\begin{pmatrix} 0 & 0 \\ 0 & 1 \end{pmatrix}, \label{YES} \\
 S_{AA} & = & e^{-i\pi/3}\begin{pmatrix} 0 & 1 \\ 1 & 0 \end{pmatrix} \label{yesss}.
\end{eqnarray}
(Deployed as a constant interlayer block these matrices will generate the standard AB/AC and AA stacked band structures.)
The matrix function $S(\br)$ can be seen to transition between AC and AB stacking with the maximum of the AA component in the middle of the partial.

\subsection{Current operator in multi-layer systems}\label{current}

In a recent article Gupta \emph{et al}.~\cite{gupta18a} have presented a derivation of the current operator in the continuum theory we employ here, finding the intuitive form

\begin{equation}
 \bj(\br) = \Im\Psi(\br)^\dagger\left[\bnab_\bp H(\br,\bp) \Psi(\br)\right]
 \label{j}
\end{equation}
with $H(\br,\bp)$ the effective Hamiltonian. 
This form, however, only describes the diagonal blocks of Eq.~\eqref{H}, where it reproduces the well known $v_F \bsig$ current operator of the Dirac-Weyl equation. We therefore require an extension of the concept of the current density to multilayer systems within the continuum approximation. 
This, however, requires some care. As each layer is represented by a strictly two-dimensional (in this case Dirac-Weyl) operator there is no natural definition of a 3-vector current density. 
The description of current flow in a multilayer therefore breaks down into an in-plane current density (a 2-vector) and an interlayer current density (a scalar), with the latter representing the net charge flow between the layers at any point $\br$. In what follows we will derive a general form of this scalar current density for a bilayer system, and show that it satisfies an appropriately generalized continuity equation.

We begin with the tight-binding form of the interlayer current operator

\begin{equation}
 \mathsf{
 j_{l}^{(z)}(\bR_i) = \frac{1}{V_{UC}} \sum_{\alpha'\in l} \left[ n(\bR_i+\bnu_{\alpha'}) v_z + v_z n(\bR_i + \bnu_{\alpha'})\right]}
\end{equation}
where the density operator is defined as $\hat{n}(\bR_i + \bnu_{\alpha'}) = \ket{\bR_i+\bnu_{\alpha'}}\bra{\bR_i+\bnu_{\alpha'}}$, the velocity operator is $v_z = 1/(i\hbar)[z,H_{TB}]$, and $V_{UC}$ is the unit-cell volume of the multilayer system
\cite{tod02,boy10}. The Hamiltonian $H_{TB}$ we take to be that of a standard two-center tight-binding approximation in a basis of Wannier orbitals, with an important caveat. We consider that the system of interest be obtained from a deformation applied to some high-symmetry reference state, for instance in this case the high-symmetry AB stacked bilayer. Note that we do not assume that this deformation is small, and thus this approach is not perturbative in nature.
We implement this deformation simply by changing the values of the tight-binding integrals while keeping the labels of the local orbitals fixed, which, evidently, represents the choice of a local co-moving coordinate system with the deformation. 
The tight-binding overlap integrals $t_{\alpha\beta}(\bR_i+\bnu_\alpha,\bR_j+\bnu_\beta)$ are then just

\begin{equation}
  t^{HS}_{\alpha\beta}(\bR_i+\bnu_\alpha,\bR_j+\bnu_\beta) +  \delta t^{LS}_{\alpha\beta}(\bR_i+\bnu_\alpha,\bR_j+\bnu_\beta) \, ,
\end{equation}
where $\alpha$ and $\beta$ are sub-lattice labels, and $\bR_i $ and $\bR_j$ lattice vectors, of the high-symmetry system. Any spin or angular momentum indices are subsumed into the $\alpha$ and $\beta$. 
The tight-binding Hamiltonian is then given by

\begin{eqnarray}
\label{tb0}
\begin{split}
H_{TB} &= \sum_{\alpha\bR_i\beta\bR_j} t_{\alpha\beta}(\bR_i+\bnu_\alpha,\bR_j+\bnu_\beta) \\  &\times c_{\bR_j+\bnu_\beta \beta}^\dagger c^{}_{\bR_i+\bnu_\alpha \alpha} \, .
\end{split}
\end{eqnarray}
The advantage of this seemingly baroque construction is that we can then naturally introduce the Bloch functions of the high-symmetry system as a complete basis for the full Hamiltonian. These are given by

\begin{equation}
 \ket{\Phi_{\bk_1\alpha}} = \frac{1}{\sqrt{N}} \sum_{\bR_i} e^{i \bk_1.(\bR_i+\bnu_\alpha)}\ket{\bR_i+\bnu_\alpha}
 \label{Bloch}
\end{equation}
with $\bk_1$ the crystal momentum and $\bnu_\alpha$ a basis vector.

We now consider a general state of the system on layer $l$ expressed in the Bloch-function basis

\begin{equation}
 \ket{\Psi_l} = \sum_{\bk_1\alpha\in l} c_{\bk_1\alpha} \ket{\Phi_{\bk_1\alpha}} \, .
\end{equation}
Using this basis expansion the interlayer current density is given by

\begin{eqnarray}
 && \mel{\Psi_l}{j_{l}^{(z)}(\bR_i)}{\Psi_{l'}} = \frac{1}{V_{UC}} \label{jz} \\
 && \times \sum_{\bk_1,\bk_2,\alpha\in l,\beta\in l'}
 \left[\mel{\Phi_{\bk_1\alpha}}{n(\bR_i+\bnu_{\alpha}) v_z}{\Phi_{\bk_2\beta}} + \text{H.c.}\ \right] \, . \nonumber
\end{eqnarray}
We now manipulate the form of the matrix element $\mel{\Phi_{\bk_1\alpha}}{n(\bR_i+\bnu_{\alpha}) v_z}{\Phi_{\bk_2\beta}}$ toward a continuum representation by introducing a hopping envelope function $t_{\alpha\beta}(\br,\bdel)$ that takes values $t_{\alpha\beta}(\bR_i+\bnu_\alpha,\bR_j+\bnu_\beta)$ for $\br = \bR_i + \bnu_\alpha$ and $\bdel = \bR_j + \bnu_\beta-\bR_i-\bnu_\alpha$. 
Inserting this function as its Fourier transform
\begin{equation}
 t_{\alpha\beta}(\br,\bdel) = \int d\bq'\,d\bq\, e^{-i\bq'.\br}e^{-i\bq.\bdel} t_{\alpha\beta}(\bq',\bq) \, ,
\end{equation}
the matrix element can be written as

\begin{eqnarray}
 &&\frac{(z_l-z_{l'})}{V_{UC} N i \hbar(2\pi)^{2d}}
 \int d\bq' d\bq\, t_{\alpha\beta}(\bq',\bq) e^{-i(\bk_1+\bq'-\bq).(\bR_i+\bnu_\alpha)} \nonumber \\
 && \times e^{i(\bk_2-\bq).\bnu_\beta} \sum_{\bR_j} e^{i(\bk_2-\bq).\bR_j}
 \label{t2}
\end{eqnarray}
with the factor $(z_l-z_{l'})$ taken outside the phase sum as all vectors in this expression are 2-vectors with no $z$-component. 
This may then be evaluated via the Poisson sum relation yielding
\begin{eqnarray}
\label{ss}
 &&\frac{1}{\sqrt{V}} e^{-\bk_1.(\bR_i+\bnu_\alpha)} \\
 &&\times \left[\frac{\text{sign}(z_l-z_l')}{i\hbar V_{UC}^\perp}
 \sum_{\bG_j}  t_{\alpha\beta}(\bR_i+\bnu_\alpha,\bk_2+\bG_j) \left[M\right]_{\alpha\beta} \right] \nonumber \\
 &&\times \frac{1}{\sqrt{V}} e^{\bk_2.(\bR_i+\bnu_\alpha)} \nonumber
 \end{eqnarray}
where we have used $(z_l-z_{l'})/V_{UC} = \text{sign}(z_l-z_l')/V_{UC}^\perp$ to relate the volume of the three-dimensional unit cell $V_{UC}$ to the in-plane unit-cell area $V_{UC}^\perp$. The plane-wave normalization factor $1/V$ arises from the $(2\pi)^2/V_{UC}^\perp$ prefactor of the Poisson sum relation along with the $1/N$ factor due to Bloch-function normalization. Introducing spinor plane waves 

\begin{figure*}
\includegraphics[width=0.90\textwidth]{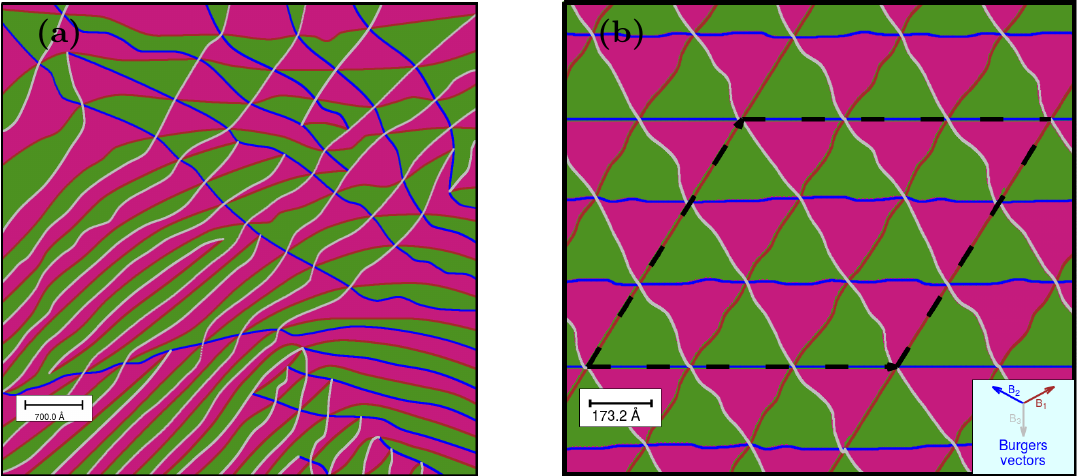}
\caption{Partial dislocation network (a) taken from digitized transmission electron microscopy images of Ref.~\onlinecite{kiss15} and (b) a  model network of trigonally intersecting partials. 
 The tile colors indicate the stacking order with red tiles AB and green tiles AC stacked. The color of the partial line indicates the Burgers vector of the partial dislocation, where the directions of the three Burgers vectors can be seen in the inset of panel (b). In the experimental network, the rectangular unit cell is reflected in the $x$- and $y$-axis to produce a periodic dislocation network. In the model network, the unit cell indicated by the dashed lines is folded out for better visual representation. 
 }
\label{networks}
\end{figure*}

\begin{equation}
\label{WCONT}
 \phi_{\bk\alpha}(\br) = \frac{1}{\sqrt{V}} e^{i\bk.\br} \ket{1_{n\alpha}} \, ,
\end{equation}
where $\ket{1_\alpha}$ is a unit ket in a space with dimensional equal the number of atomic degrees of freedom, \emph{i.e.}\ $\ket{1_{\alpha}} = (0_{1},\ldots,1_{\alpha},\ldots)^T$, we then find for the tight-binding matrix element the final result

\begin{equation}
 \frac{\text {sign}(z_l-z_l')}{i\hbar}
 \phi_{\bk_1\alpha}^\dagger(\br) H(\bR_i+\bnu_\alpha,\bp) \phi_{\bk_2\beta}(\br) \nonumber
\end{equation}
where, by writing the Bloch-state momentum $\bk_2$ relative to an expansion point $\bK_0$ \emph{i.e.}\ substituting $\bG_j+\bk_2 = \bG_j + \bK_0 + \bp_2$ in Eq.~\eqref{ss}, and promoting $\bp_2$ to an operator, we have identified

\begin{equation}
 H(\br,\bp) = \frac{1}{V_{UC}^\perp}
 \sum_{\bG_j} \left[M\right]_{\alpha\beta} t_{\alpha\beta}(\br,\bK_j + \bp)
\end{equation}
as the effective continuum Hamiltonian described in Ref.~\onlinecite{M} from which Eq.~\eqref{S} may be obtained.

Using Eq.~\eqref{jz} immediately yields for the $z$-component of current in layer 1

\begin{equation}
j_{1}^{(z)}(\br) = \frac{1}{i\hbar}\left[-\Psi_1^\dagger (H_{12} \Psi_2) + ( H_{12} \Psi_2)^\dagger \Psi_1\right]
\label{jzf}
\end{equation}
where we take the origin of the $z$-coordinate to be within layer 1 pointing toward layer 2. 
This result is unsurprising, being identical in form to the standard ``bond-current'' expression\cite{tod02}. However, rather than a current along a bond it represents a current between layers whose coupling is through the continuum field $H_{12}(\br)$.
The use of such a local coordinate system ensures that, due to the factor $\text{sign}(z_l-z_{l'})$, we have
\begin{equation}
 j_{1}^{(z)}(\br) = -j_{2}^{(z)}(\br) \, .
\end{equation}
We now demonstrate that this form of the interlayer current satisfies the appropriate generalization of the continuity equation. To that end we consider the time dependence of the density within layer 1

\begin{eqnarray}
 \partial_t n_1(\br) & = & \frac{1}{i\hbar}\left[\Psi_1^\dagger(H_{11}\Psi_1)-(H_{11}\Psi_1)^\dagger\Psi_1\right] \nonumber \\
 &+&\frac{1}{i\hbar}\left[\Psi_1^\dagger(H_{12}\Psi_2)-(H_{12}\Psi_2)^\dagger\Psi_1\right] \, ,
 \label{contz0}
\end{eqnarray}
where we have used

\begin{equation}
 \begin{pmatrix} H_{11} & H_{12} \\ H_{21} & H_{22} \end{pmatrix} \begin{pmatrix} \Psi_1 \\ \Psi_2 \end{pmatrix} =
 i\hbar \partial_t \begin{pmatrix} \Psi_1 \\ \Psi_2 \end{pmatrix}
\end{equation}
in order to replace time derivatives. 
The first term on the right hand side of Eq.~\eqref{contz0} is just the negative of the divergence of intralayer current, with the second term just the negative of $j^{(z)}_1$, as may be seen from Eq.~\eqref{jzf}. 
We then immediately find a generalized continuity equation given by

\begin{equation}
 \partial_t n_1(\br) + \bnab.\bj(\br) + j_{1}^{(z)}(\br) = 0
\end{equation}
with a corresponding relation for layer 2. 
If $\Psi(\br)$ is a stationary state of the system then

\begin{equation}
 \int d\br j_{l}^{(z)}(\br) = 0
\end{equation}
which follows from the Hermiticity of the layer diagonal blocks $H_{ll}$.

\section{Partial dislocation networks}

\subsection{Numerical details}

We employ as a basis the Dirac-Weyl states of each layer, which has the advantage of compactness as the energy range required for convergence is $\sim 1.5$ the energy window of interest. 
The interlayer stacking potential $S(\br)$ is defined on a fine grid (we use $3000\times3000$ for the experimental network and $800\times800$ for the model network) and the matrix elements of $S(\br)$ in the Dirac-Weyl basis obtained via FFT. For the large networks that we consider, a basis of up to 25,000 Dirac-Weyl states from each layer is required, necessitating a distributed-memory parallelization approach to diagonalization. For the interlayer tight-binding hopping function we use a Gaussian form $t_\perp(\bdel^2) = A \exp(-B\bdel^2)$, with $A = 0.4\,$eV and $B = 1\,$\si{\angstrom}$^{-2}$ chosen to reproduce the $0.8\,$eV splitting of bonding and anti-bonding bands in the Bernal bilayer.

\begin{figure*}
  \centering
  \includegraphics[width=0.85\linewidth]{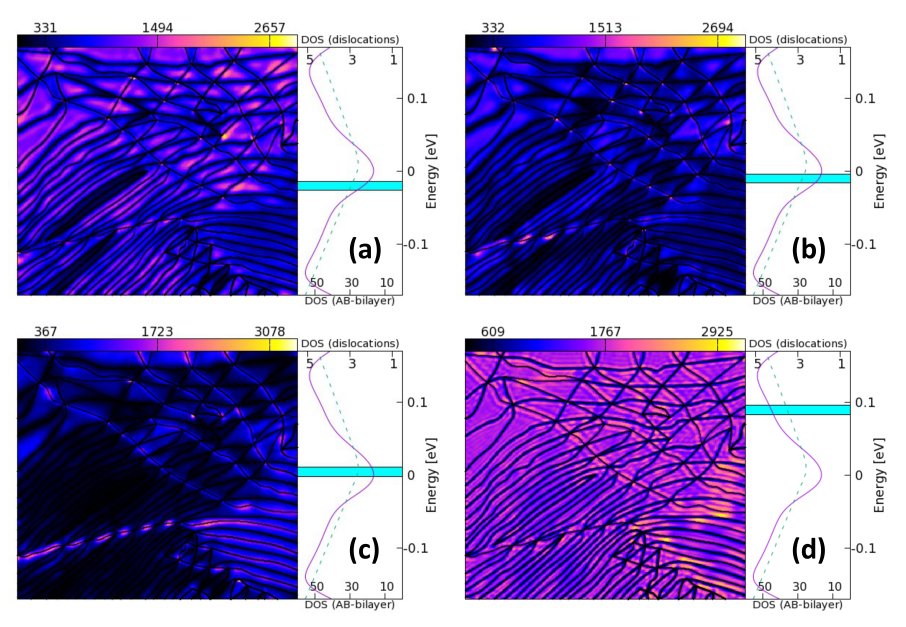}
  \caption{Electron density obtained by integrating over all states within a 13~meV energy window situated at 4 different energies in the density of states (DOS) of the partial dislocation network: -20~meV (a), -5~meV (b), +5~meV (c), and +90~meV (d). Close to the Dirac point, panels (b) and (c), charge pooling on the segments of the mosaic network can be observed. In panel (a) charge localization associated with partials of type 3 can be seen which, above the Dirac point energy, switches to a charge accumulation associated with partials of type 2, see panels (c) and (d). The right hand panel in each case indicates the energy window within which states are summed to give the electron density, with the full line the network DOS and the dashed line the Bernal bilayer DOS presented for comparison.
  }
  \label{expd}
\end{figure*}

\begin{figure*}
  \centering
  \includegraphics[width=0.98\linewidth]{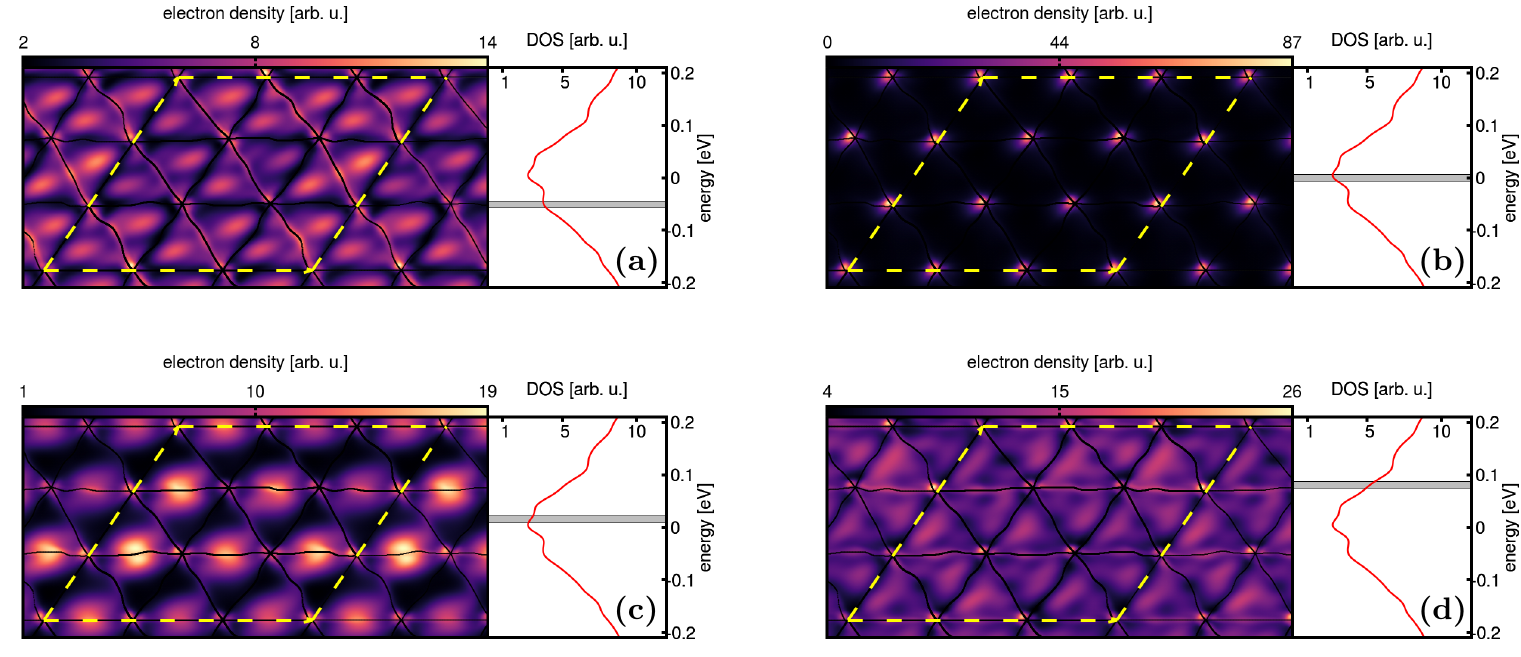}
  \caption{Electron density obtained by integrating all states over a 13~meV energy window situated at 4 different energies: (a) -55~meV, (b) 0~meV, (c) +17~meV, and (d) +87~meV. We find localization on partials with the same energy ordering as may be observed in the more complex experimentally derived partial network (Fig.~\ref{expd}): type 3 below the Dirac point as seen in panel (a), and type 2 above the Dirac point as seen in panel (c). In contrast to the experimental network, localization on the nodes of the network close to the Dirac point in now very pronounced, see panel (b).
    }
  \label{hexd}
\end{figure*}

\begin{figure*}
  \centering
  \includegraphics[width=0.98\linewidth]{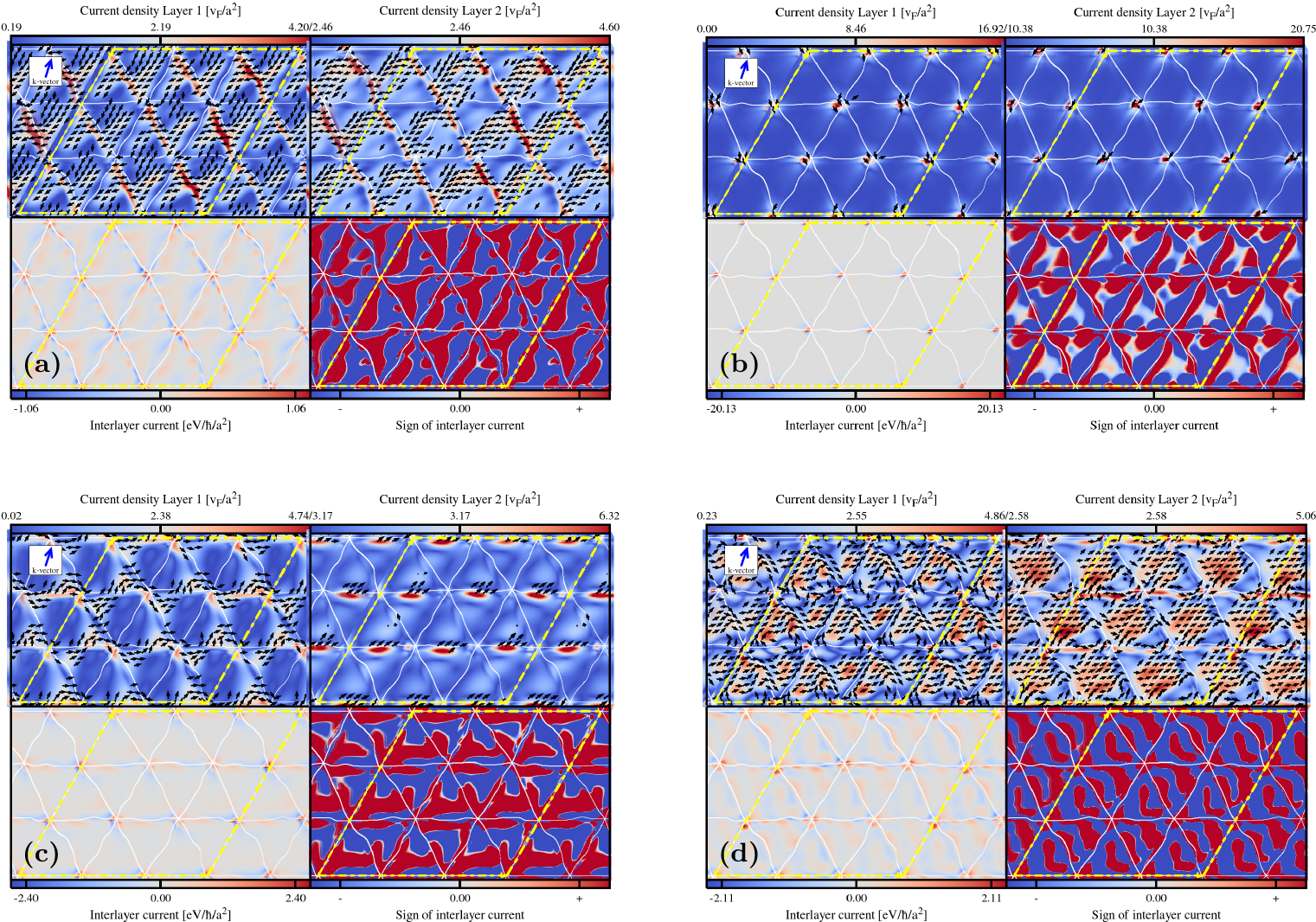}
  \caption{Interlayer and intralayer currents for the network states whose electron density is exhibited in Fig.~\ref{hexd}. In each panel are shown the current density in layer 1, the current density in layer 2, the interlayer current density, and for clarity the sign of the interlayer current density. The charge accumulation states are seen to be current carrying, especially clear in panel (c) in which the current density is seen to ``spiral'' around the partial. In all panels it is seen that the interlayer current is pronounced at partial dislocations.
  }
  \label{hexj}
\end{figure*}

\subsection{Charge pooling on the mosaic network and charge accumulation at partials}

We will first consider the electronic structure of a partial dislocation network taken directly from transmission electron microscopy (TEM) images\cite{kiss15} of bilayer graphene grown epitaxially on SiC. 
This network is illustrated in Fig.~\ref{networks} with the various partial dislocations colored according to the their Burgers vector. 
Exploratory calculations of this system have already been presented\cite{kiss15}, here we will investigate in greater detail the electronic structure. 
The area of the TEM image in experiment was 1$\mu m^2$ equivalent to $\approx\!10^8$ carbon atoms which, within the continuum approach described in the previous section, is numerically feasible. 
The interlayer stacking potential $S(\br)$, Eq.~\eqref{S}, that depends directly on the local translation $\Delta \bu(\br)$, must now encode the mutual translation of the layers that occurs upon crossing any partial line in the network (within the domains of the mosaic structure the function $\Delta \bu(\br)$ will obviously be constant). 
The partial dislocations in Fig.~\ref{networks} evidently have a mixture of edge and screw dislocation character,  corresponding to the cases in which the Burgers vector is, respectively, parallel or perpendicular to the tangent of the partial line. 
To model this we take the form of the interlayer displacement Eq.~\eqref{disp} shown in Fig.~\ref{uphi} and use this to describe the displacement along the line perpendicular to the partial tangent, forming a superposition of the resulting displacements at each point in space as this procedure is traced out along a partial line. In this way we find a smooth interlayer displacement field describing a network of wandering partials. 
Note that a local shift by a multiple of the lattice vectors $\ba_{1,2}$ does not alter the interlayer potential, due to the presence of the reciprocal lattice vector in the exponent of Eq.~\eqref{S}. 
%Thus the relation between the deformation field and the interlayer potential is, in principle, multi-valued.

Of principle interest is the form of the wave functions of the partial network, which are expected to be very different from the wave functions of the structurally perfect 
bilayer. 
In Fig.~\ref{expd}(a-d) we show the electron density integrated over a 13~meV 
window (of the order of the Fermi smearing at 150 K) with this window placed at four different energies: -20~meV, -5~meV, +5~meV, +90~meV. 
Even within this small energy window, of the order of $10^3$ individual eigenstates contribute to the probability density. 
Quite clearly, the mosaic structure of the bilayer has a dramatic impact on the wave functions. For the density integrated in the energy window situated at -20meV, see panel (a) of Fig.~\ref{expd}, charge accumulation is found on the type 3 partials (compare with Fig.~\ref{networks}) but not on the partials of type 1 and 2. 
Above the Dirac point this switches to a charge accumulation on type 2 partials, see panels (c) and (d). 
Close to the Dirac point substantial charge pooling is observed as some segments of the mosaic network have significantly higher density than others (see panels (b) and (c)), a point first noticed in Ref.~\onlinecite{kiss15}. Intense localization on intersections of the partial network can also be noted. For the energy window of +90meV, shown in panel (d), the pronounced charge pooling seen near the Dirac point is absent, although one still notes substantial density inhomogeneity created by the partial network. 

Now turning to the features of the density of states (DOS), we note qualitative differences between pristine AB-stacked graphene bilayer and that of the bilayer with the experimentally observed partial dislocation network. At the Dirac point, the experimental system features a much more pronounced minimum of the DOS, with two side peaks found at approximately $\pm0.14$~eV.

How much of this structure, and in particular the existence of ``hot'' partials on which 
charge density is accumulated, is related to the specific partial network shown in Fig.~\ref{expd}? In this respect it should be noted that charge accumulation on partials occurs on differently oriented segments of a wandering partial, and so the local orientation of the partial with respect to the underlying lattice is evidently only of secondary importance. This is particularly the case for the type 2 partial, as may be seen in panel (c) of Fig.~\ref{expd}.

To investigate further these features we consider a designed hexagonal network of partial dislocations shown in Fig.~\ref{networks}. 
In this network all partials are globally aligned along the zigzag direction of the underlying lattice, however we introduce some random disorder into the partials such that they are not perfectly straight, and so locally are not necessarily aligned with the zigzag direction.

As may be observed in Fig.~\ref{hexd}, qualitatively similar features are seen to those noted in the experimental partial network - in particular the energy order in which the type 3 and 2 partials become "hot" and accumulate charge is the same; the type 1 partial, as before, does not accumulate charge. The width of the energy window is 
again $13\,$meV and the position of the energy windows similar to those used in the analysis of the experimental network. This charge accumulation is therefore largely independent of the global details of partial network but sensitive to energy and Burgers vector type. A number of features are much more pronounced in this more ordered partial network, in particular the localization of charge on the nodes, see panel (d) of Fig.~\ref{hexd}. 
A complete sweep of the electron density as a function of energy can be obtained from the electronic supplementary information.
% node states found in what energy window?
% Energy window for node states and other features.

\subsection{Current density in the partial network}

The strong imprint of the partial network upon electron density raises the natural question as to the corresponding current densities. Using the formalism outlined in Section \ref{current} we now examine this question. In Fig.~\ref{hexj} we present the current density for those states whose electron density is exhibited in Fig.~\ref{hexd}. In each panel is shown the in-plane current for each layer and the interlayer current density. For clarity we also show the sign of the interlayer current. All current densities shown here are, of course, exactly compensated by a corresponding state in the K$^\ast$ valley with opposite current.

We first investigate the current density of the charge accumulation states shown in panels (a) and (c) of Fig.~\ref{hexd}. 
As can be seen in the corresponding panels of Fig.~\ref{hexj} both these states are current carrying. This is particularly evident in panel (c) for the type 2 partial dislocation: the interlayer current exhibits an alternating sign along the partial dislocation line that, in conjunction with an in-plane current density alternating between the layers, signifies that the current spirals around the partial, from one layer to the other and back again. The type 2 charge accumulation state thus appears similar to the topological current-carrying state found in the gap generated by application of finite interlayer bias. The results here raise the intriguing possibility of current-carrying partial dislocation states in the absence of such an external field. This can also be seen in panel (a) for the type 3 partial, although the structure is somewhat obscured.

In general we find that the partial dislocations are associated with strong interlayer currents, as may be seen from panels (a)-(d) of Fig.~\ref{hexj}. It is evident by comparison of the in-plane currents that the interlayer current is intense near partials. This can be seen for example in panel (a) upon examination of the charge repelling type 1 partial. One notes that flow of charge in layer 1 appears to be interrupted by the partial. However, comparison with the current density in layer 2 and the interlayer current shows that the charge simply tunnels underneath the dislocation. This finding of strong interlayer currents associated with partial dislocations is consonant with the situation found in the other common stacking fault of bilayer graphene, a mutual rotation of the layers. In that case snake states are found that oscillate between the two layers, and the spiral state found on the type 2 partial can be seen as the analogy of this. Evidently, in bilayer graphene stacking modulation generate strong interlayer current densities.

Finally, we consider the situation of the nodal states and the high-energy states that show no pronounced pattern of charge accumulation. In panels (b) and (d) of Fig.~\ref{hexj} we show the current density corresponding to the electron densities shown in panels (b) and (d) of Fig.~\ref{hexd}. The nodal states are seen to exhibit intense current patterns in both the in-plane and interlayer currents. Evidently, nodal states are associated with interlayer current loops, quite tightly bound to the nodes. 
The current density does not, however, exhibit a six-fold pattern corresponding to the basic geometry of the intersection point -- the low-symmetry environment of the node due to partial wandering precludes this. 
For high-energy states, that are neither nodal nor partial dislocation states, the current density is correspondingly featureless. 
One does however note that the strong interlayer current persists and again generally changes sign across a partial dislocation line.

%%%%%%%%%%%%%%%
% CONCLUSIONS %
%%%%%%%%%%%%%%%

\section{Discussion}

We have explored the electronic structure of massive realistic partial dislocation networks featuring wandering partials that both intersect with each other and annihilate in pairs at local defects. Strong charge accumulation at partials is found that is sensitive to energy and Burgers vector but less sensitive to the orientation of the partial line with respect to the underlying lattice or to the screw versus edge character of the partial. 
These charge accumulation states are all found within $30\,$meV of the 
Dirac point and are current-carrying with the current density exhibiting a spiral motion around the partial line. 
We find that partial dislocations are generally associated with strong interlayer current flow, a feature found not only for charge accumulation states but also in states for which intralayer current density is blocked by a partial but ``short circuits'' by flowing into the other layer. 
Very close to the Dirac point, charge strongly localizes on the nodes of the partial dislocation network, in particular for more ordered networks but less so for disordered partial networks typically found in experiment. 
Correspondingly, the current density exhibits a complex current flow around the nodes.

While current-carrying topological states are known to be found in the band gap of a bilayer subjected to interlayer bias, these states require no external bias for their existence. Partial dislocations in realistic networks are thus natural lines for the flow of current density, with potential implications for the quantum Hall effect in mosaic bilayer graphene.

\bibliographystyle{unsrt}
\bibliography{disloc}

\end{document}